\documentclass[apjl]{emulateapj}
\usepackage{mathptmx}

\newcommand{\ltsim}{\raisebox{-.5ex}{$\;\stackrel{<}{\sim}\;$}}
\newcommand{\gtsim}{\raisebox{-.5ex}{$\;\stackrel{>}{\sim}\;$}}
\newcommand{\kms}{\ifmmode {\rm km\ s}^{-1} \else km s$^{-1}$\fi}
\newcommand{\vFWHM}{\ifmmode V_{\mbox{\tiny FWHM}} \else
            $V_{\mbox{\tiny FWHM}}$\fi}
\newcommand{\msun}{$M_{\odot}$}
\newcommand{\et}{et al.\ }

\newcommand{\hb}{H$\beta$}
\newcommand{\mbh}{$M_{\rm BH}$}
\newcommand{\lledd}{$L/L_{\rm Edd}$}
\newcommand{\lbol}{$L_{\rm bol}$}
\newcommand{\aox}{$\alpha_{\rm ox}$}
\newcommand{\nh}{$N_{\rm H}$}
\newcommand{\Ka}{\hbox{Fe K$\alpha$}}
\newcommand{\xmm}{{\hbox{\sl XMM-Newton}}}

\newcommand{\xray}{\hbox{X-ray}}

\newcommand{\qthirteenlong}{\hbox{Q\,1346$-$036}}
\newcommand{\hetwotwo}{\hbox{HE\,2217$-$2818}}

\journalinfo{The Astrophysical Journal, 
???:L??--L??, 2006 ??? ??, astro-ph/0606389}
\slugcomment{Received 2006 May 11; accepted 2006 June 14}

\shorttitle{X-RAY SPECTRAL SLOPE--ACCRETION RATE RELATIONSHIP IN AGNs}
\shortauthors{SHEMMER ET AL.}

\begin{document}

\title{The Hard X-ray Spectral Slope as an Accretion-Rate Indicator in
  Radio-Quiet \\ Active Galactic Nuclei}

\author{
Ohad Shemmer,\altaffilmark{1}
W.~N. Brandt,\altaffilmark{1}
Hagai Netzer,\altaffilmark{2}
Roberto Maiolino,\altaffilmark{3}
and Shai Kaspi\altaffilmark{2,4}
}

\altaffiltext{1} {Department of Astronomy \& Astrophysics,
  Pennsylvania State University, University Park, PA 16802, USA;
  ohad@astro.psu.edu.}

\altaffiltext{2} {School of Physics \& Astronomy, Raymond and Beverly
  Sackler Faculty of Exact Sciences, Tel Aviv University, Tel Aviv
  69978, Israel.}

\altaffiltext{3} {INAF - Osservatorio Astrofisico di Arcetri, L.go
  E. Fermi 5, 50125 Firenze, Italy.}

\altaffiltext{4} {Physics Department, Technion, Haifa 32000, Israel.}

\begin{abstract}
We present new \xmm\ observations of two luminous and high
accretion-rate radio-quiet active galactic nuclei (AGNs) at
$z\sim2$. Together with archival \xray\ and rest-frame optical spectra
of three sources with similar properties as well as 25
moderate-luminosity radio-quiet AGNs at $z<0.5$, we investigate, for
the first time, the dependence of the hard (\gtsim2\,keV)
\xray\ power-law photon index on the broad \hb\ emission-line width
and on the accretion rate across $\sim3$ orders of magnitude in AGN
luminosity. Provided the accretion rates of the five luminous sources
can be estimated by extrapolating the well-known broad-line region
size-luminosity relation to high luminosities, we find that the photon
indices of these sources, while consistent with those expected from
their accretion rates, are significantly higher than expected from the
widths of their \hb\ lines. We argue that, within the limits of our
sample, the hard-\xray\ photon index depends primarily on the
accretion rate.
\end{abstract}

\keywords{galaxies: active -- galaxies: nuclei -- X-rays: galaxies --
  quasars: emission lines -- quasars: individual (\qthirteenlong,
  \hetwotwo)}

\section{What Determines the X-ray Power-Law Spectrum of Active Galactic
Nuclei?}
\label{introduction}

It is widely accepted that the hard (\gtsim2\,keV) \xray\ emission
from active galactic nuclei (AGNs) is primarily produced via
unsaturated inverse Compton scattering of UV--soft-\xray\ photons from
the accretion disk by a corona of hot, likely thermal, electrons
(e.g., Haardt \& Maraschi 1991; Zdziarski, Poutanen, \& Johnson 2000;
Kawaguchi, Shimura, \& Mineshige 2001). The emitted \xray\ photon
spectrum in the \hbox{$\approx$2--100\,keV} energy range is best
described by a power-law of the form $N_E\propto E^{-\Gamma}$, where
$\Gamma$, the photon index, takes a fairly constant value of $\sim$2,
and is predicted to be only weakly sensitive to large changes in the
electron temperature and the optical depth in the corona (e.g., Haardt
\& Maraschi 1991). Throughout this work, we refer to $\Gamma$ only as
the photon index in the hard (\gtsim2\,keV) \xray\ band.

Nandra \& Pounds (1994) found a $\left < \Gamma \right
>$=1.95$\pm$0.15 for a sample of bright Seyfert\,1 galaxies,
confirming the mean predicted $\Gamma$ and its small dispersion.
Brandt, Mathur, \& Elvis (1997) have studied the \xray\ spectra of a
sample of nearby AGNs, including low--moderate luminosity Narrow-Line
Seyfert~1 (NLS1) galaxies, defined as type~1 AGNs having \hb\ full
width at half-maximum intensity (FWHM)\ltsim2000\,\kms\ (Osterbrock \&
Pogge 1985). The addition of NLS1s, which were previously overlooked
due to observational biases, broadened the range of $\Gamma$ values,
and allowed Brandt \et (1997) to find that $\Gamma$ is anticorrelated
with FWHM(\hb). This extended the analogous relation between FWHM(\hb)
and the effective \xray\ spectral slope in the \hbox{0.1--2.0\,keV}
band to harder \hbox{X-rays} (e.g., Boller, Brandt, \& Fink 1996;
Laor~\et~1997). The pseudo-thermal soft-\xray\ excess radiation may be
responsible for cooling the corona, thus steepening the
\xray\ spectrum (e.g., Pounds, Done, \& Osborne 1995), and the
pronounced soft excesses of NLS1s (e.g., Puchnarewicz \et 1995) may
readily explain their steep \xray\ spectra and the $\Gamma$-FWHM(\hb)
anticorrelation.

The remarkable dependence of the spectral shape of the
\xray\ emission, originating\ltsim30\,$R_{\rm S}$ from the central
engine, on the width of a broad-emission line region (BELR) line,
emitted at $\approx$10$^{4}\,R_{\rm S}$, was interpreted as a
fundamental dependence of $\Gamma$ on the accretion rate, since, as
explained below, FWHM(\hb) is considered an accretion-rate indicator
in type~1 AGNs (e.g., Boroson \& Green 1992; Brandt \& Boller 1998). A
high accretion rate would increase the disk temperature hence
producing more soft-\xray\ radiation (the ``soft excess'') and, at the
same time, increase the Compton cooling of the corona and steepen
the hard-\xray\ power law.

Recent \xray\ studies of nearby ($z$\ltsim0.5) type~1 radio-quiet
quasars (RQQs) have consistently shown that $\Gamma \sim2.0\pm0.5$,
and some of those studies confirmed the Brandt \et (1997)
$\Gamma$-FWHM(\hb) relationship (e.g., Leighly 1999; Reeves \& Turner
2000; Porquet \et 2004; Piconcelli \et 2005, hereafter P05; Brocksopp
\et 2006). Photon indices of $\Gamma \sim2.0$ are also observed for
0.5\ltsim $z$\ltsim6 RQQs (e.g., Reeves \& Turner 2000; Page \et 2005;
Shemmer \et 2005), however, the spread in their $\Gamma$ values seems
smaller than that for nearby sources (see Fig.\,3 of Shemmer \et
2005). On the other hand, Grupe \et (2006) report $\left < \Gamma
\right > =2.21\pm0.52$ for a sample of $z>4$ RQQs, and suggest that
the relatively steep \xray\ spectra of such sources may be attributed
to high accretion rates, in analogy with NLS1s.

\begin{deluxetable*}{lcccccccc}
\tablecolumns{9}
\tabletypesize{\scriptsize}
\tablewidth{0pc}
\tablecaption{{\sl XMM-Newton} Observation Log \label{obs_log}}
\tablehead
{
\colhead{} &
\colhead{} &
\colhead{} &
\colhead{} &
\colhead{} &
\colhead{{\sc Observation}} &
\multicolumn{3}{c}{{\sc Net Exposure Time (ks) / Source Counts}} \\
\colhead{{\sc Quasar}} &
\colhead{{\sc RA (J2000.0)}} &
\colhead{{\sc DEC (J2000.0)}} &
\colhead{$z$\tablenotemark{a}} &
\colhead{$N_{\rm H}$\tablenotemark{b}} &
\colhead{{\sc Start Date}} &
\colhead{MOS1} &
\colhead{MOS2} &
\colhead{{\sc pn}}
}
\startdata
\qthirteenlong\ & 13 48 44.08 & $-$03 53 25.0 & 2.370 &
2.47 & 2005 Jul 05 & 10.4 / 90 & 10.4 / 144 & 9.4 / 520 \\
\hetwotwo\ & 22 20 06.77 & $-$28 03 23.9 & 2.414 &
1.28 & 2005 Oct 31 & 25.1 / 732 & 25.1 / 790 & 21.2 / 2968
\enddata
\tablenotetext{a}{Systemic redshift measured from the optical emission
  lines and obtained from S04.}
\tablenotetext{b}{Neutral Galactic absorption column density in units
  of $10^{20}$\,cm$^{-2}$ obtained from Dickey \& Lockman (1990).}
\end{deluxetable*}

In this {\it Letter} we test the hypothesis that the accretion rate
largely determines the hard-\xray\ spectral slope in AGNs, and argue
that previous studies have not been able to break the degeneracy
between the dependence of $\Gamma$ on FWHM(\hb) and the accretion
rate, since highly luminous AGNs were left out of the analyses (e.g.,
Porquet \et 2004). This is performed by investigating \xray\ and
\hb\ spectral-region data for a well-defined sample of 30
moderate--high-luminosity RQQs, selected for having high-quality
\xray\ and \hbox{\hb-region} spectroscopy. Five of the sources are
luminous RQQs at $z\sim2$, allowing, for the first time in this
context, expansion of the AGN parameter space by $\sim3$ orders of
magnitude in luminosity and black-hole mass (\mbh). Two of the
$z\sim2$ RQQs, \qthirteenlong\ and \hetwotwo, were selected for
\xmm\ observations from the Shemmer \et (2004; hereafter S04) sample
of luminous and high accretion-rate quasars based on their expected
high \xray\ fluxes in that sample. In \S\,\ref{observations} we
present the new \xmm\ observations, their reduction, and the
\xray\ spectral fitting of these two sources; their \xray\ and optical
properties are presented in \S\,\ref{individual}. In
\S\,\ref{indicator} we discuss our results, and in particular, the
dependence of $\Gamma$ on the accretion rate for AGNs. Throughout this
work we consider only radio-quiet AGNs to avoid any contribution from
jet-related emission to the \xray\ spectra. Luminosity distances are
computed using the standard cosmological model with parameters
$\Omega_{\Lambda}=0.7$, $\Omega_{M}=0.3$, and
$H_{0}=70$\,\kms\,Mpc$^{-1}$.

\section{XMM-{\em Newton} Observations and Data Reduction}
\label{observations}

\begin{figure}
\centering
\plotone{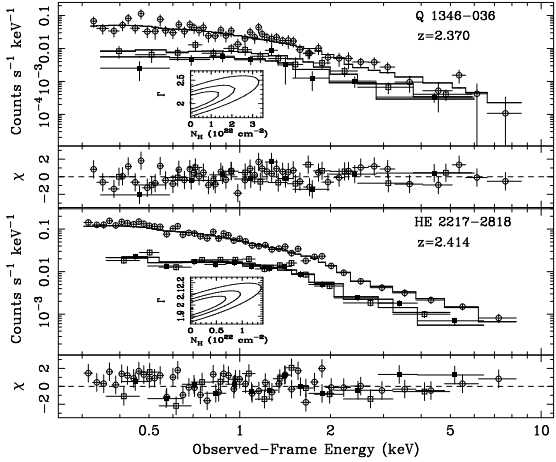}
\caption{Data, best-fit spectra, and residuals for the
  \xmm\ observations of \qthirteenlong\ ({\it top}) and
  \hetwotwo\ ({\it bottom}). Open circles, filled squares, and open
  squares represent the EPIC pn, MOS1, and MOS2 data,
  respectively. Solid lines represent the best-fit model for each
  spectrum, and the thick line marks the best-fit model for the pn
  data. The data were fitted with a Galactic-absorbed power-law model
  in the observed-frame $>$0.6\,keV band, that was extrapolated to
  0.3\,keV in the observed frame. The $\chi$ residuals are in units of
  $\sigma$ with error bars of size 1. The inset in each panel shows
  68\%, 90\%, and 99\% confidence contours for $\Gamma$ and \nh, when
  the data are fitted with an additional neutral intrinsic-absorption
  component.}
\label{spectra}
\end{figure}

Table\,\ref{obs_log} gives a log of the \xmm\ (Jansen \et 2001)
imaging spectroscopic observations for \qthirteenlong\ and \hetwotwo;
the data were processed using standard \xmm\ Science Analysis
System\footnote{http://xmm.esac.esa.int/sas.} v6.5.0 tasks. The event
files of the observation of \qthirteenlong\ were filtered to remove
$\sim$35\,ks of flaring-activity periods; the net exposure times in
Table\,\ref{obs_log} reflect the filtered data (time filtering was not
required for the \hetwotwo\ observation). The \xray\ spectra of the
quasars were extracted from the images of all three European Photon
Imaging Camera (EPIC) detectors using apertures with radii of
30\arcsec. Background regions were at least as large as the source
regions. The spectra of \qthirteenlong\ (\hetwotwo) were grouped with
a minimum of 10 (50) counts per bin. Joint spectral fitting of the
data from all three EPIC detectors for each quasar was performed with
{\sc xspec} v11.3.2 (Arnaud 1996). We employed Galactic-absorbed
power-law models at observed-frame energies $>$0.6\,keV, corresponding
to $>$2\,keV in the rest-frame of each quasar, where the underlying
power-law hard-\xray\ spectrum is less prone to contamination due to
any potential soft excess emission and ionized absorption. The
best-fit $\Gamma$ values and power-law normalizations from these fits
are given in Table\,\ref{properties}, and the data, their joint,
best-fit spectra, and residuals appear in Fig.\,\ref{spectra}.

We searched for intrinsic absorption in each quasar by jointly fitting
the spectra with a Galactic-absorbed power law model including an
intrinsic (redshifted) neutral-absorption component with solar
abundances in the same energy range as before. None of the quasars
shows significant intrinsic absorption; upper limits on intrinsic
\nh\ values appear in Table\,\ref{properties}. Fig.\,\ref{spectra}
includes a \hbox{$\Gamma$-\nh} confidence-contour plot from this
fitting for each quasar. By applying $F$-tests between the models
including intrinsic absorption and those that exclude it, we found
that neither data set requires an intrinsic absorption component. We
also note that our \xmm\ spectra show no indication of
Compton-reflection components or \Ka\ lines; such features are
expected to be relatively weak, below our detection threshold, in the
luminous sources under study in this work (e.g., the \Ka\ equivalent
width is expected to be\ltsim50\,eV; Page \et 2004a).

Finally, we searched for rapid ($\sim$1\,hr timescale in the rest
frame) \xray\ variations in the \xmm\ data of the two quasars by
applying Kolmogorov-Smirnov (KS) tests to the lists of photon arrival
times from the event files, but no significant variations were
detected.

\begin{deluxetable*}{lccccccccc}
\tablecolumns{10}
\tabletypesize{\scriptsize}
\tablewidth{0pc}
\tablecaption{Best-Fit X-Ray Spectral Parameters and Optical
  Properties \label{properties}}
\tablehead{
\colhead{} &
\colhead{} &
\colhead{} &
\colhead{} &
\colhead{} &
\colhead{$\log \nu L_{\nu}(5100\,\mbox{\AA})$} &
\colhead{FWHM(\hb)} &
\colhead{$\log M_{\rm BH}$} &
\colhead{} &
\colhead{} \\
\colhead{{\sc Quasar}} &
\colhead{{\sc $\Gamma$}} &
\colhead{$f_{\nu}$(1\,keV)\tablenotemark{a}} &
\colhead{$\chi^{2}$~(d.o.f)} &
\colhead{\nh\tablenotemark{b}} &
\colhead{(ergs\,s$^{-1}$)} &
\colhead{(\kms)} &
\colhead{(\msun)} &
\colhead{\lledd} &
\colhead{\aox} \\
\colhead{(1)} &
\colhead{(2)} &
\colhead{(3)} &
\colhead{(4)} &
\colhead{(5)} &
\colhead{(6)} &
\colhead{(7)} &
\colhead{(8)} &
\colhead{(9)} &
\colhead{(10)}
}
\startdata
\qthirteenlong\ & $2.02\pm0.17$ & $1.6\pm0.2$ & 37~(52) & $\le1.2$ &
46.9 & 5100 & 10.0 & 0.3 &
$-$1.69 \\
\hetwotwo\ & $1.97\pm0.06$ & $3.5\pm0.2$ & 50~(50) & $\le0.6$ & 47.2 &
5200 & 10.3 & 0.3 &
\ \ \ $-$1.69
\enddata
\tablecomments{The best-fit photon index, normalization, and $\chi^2$
  were obtained from a Galactic-absorbed power-law model. Errors
  represent 90\% confidence limits, taking one parameter of interest
  ($\Delta \chi^{2}=2.71$).}
\tablenotetext{a}{Power-law normalization for the pn data, taken from
  joint fitting of all three EPIC detectors with the Galactic-absorbed
  power-law model, and is given as the flux density at an
  observed-frame energy of 1\,keV with units of
  10$^{-31}$\,ergs\,cm$^{-2}$\,s$^{-1}$\,Hz$^{-1}$.}
\tablenotetext{b}{Intrinsic column density in units of
  10$^{22}$\,cm$^{-2}$. Upper limits were computed with the
  intrinsically-absorbed power-law model with Galactic absorption, and
  represent 90\% confidence limits for each value.}
\end{deluxetable*}

\section{X-Ray and Optical Properties}
\label{individual}

Basic \xray\ and optical properties of \qthirteenlong\ and
\hetwotwo\ are given in Table\,\ref{properties}. The luminosity at a
rest-frame wavelength of 5100\,\AA\ is given in column\,(6), and
FWHM(\hb) is given in column\,(7); these data were obtained from
S04. Column\,(8) gives \mbh, determined as

\begin{equation}
\label{eq:mbh}
\frac{M_{\rm BH}}{10^{6}M_{\sun}}=4.35\left[\frac{\nu
    L_{\nu}(5100\,\mbox{\AA})}{10^{44}\,{\rm
      ergs\,s^{-1}}}\right]^{0.7}\left[\frac{{\rm FWHM}({\rm
      H}\beta)}{10^3\,{\rm km\,s^{-1}}}\right]^2,
\end{equation}
and based on the recent reverberation-mapping results of Peterson \et
(2004) and the Kaspi \et (2005) BELR size-luminosity relation (see
also Kaspi \et 2000). We note that this relation relies on a sample of
AGNs having luminosities up to $\approx$10$^{46}$\,ergs\,s$^{-1}$, and
extrapolating it to higher luminosities is somewhat uncertain; a
reverberation-mapping effort is underway to test the validity of such
extrapolations (S.~Kaspi et al., in preparation). However, as \mbh\ is
expected to scale with luminosity, for a given \hb\ width, and since
in this work we perform only nonparametric statistical ranking tests,
our results are not significantly sensitive to the precise coefficient
values in this relation. Using equation (\ref{eq:mbh}), the
accretion-rate ratios (col. [9]), \lbol/$L_{\rm Edd}$ (where \lbol\ is
the bolometric luminosity; hereafter \lledd), are given by

\begin{equation}
\label{eq:lledd}
L/L_{\rm Edd}=0.15f(L)\left[\frac{\nu
    L_{\nu}(5100\,\mbox{\AA})}{10^{44}\,{\rm
      ergs\,s^{-1}}}\right]^{0.3}\left[\frac{{\rm FWHM}({\rm
      H}\beta)}{10^3\,{\rm km\,s^{-1}}}\right]^{-2},
\end{equation}
where we have employed equation (21) of Marconi \et (2004) to obtain
$f(L)$, which is the luminosity-dependent bolometric correction to
$\nu L_{\nu}(5100\,\mbox{\AA})$. The bolometric correction is
\hbox{$f\sim6-8$} \hbox{($f\simeq5$)} for the $z<0.5$ ($z\sim2$)
sources in this work.

The optical--\xray\ spectral slopes in column\,(10) are defined as
\aox$=\log(f_{\rm 2\,keV}/f_{2500\mbox{\rm\,\scriptsize\AA}})/
\log(\nu_{\rm 2\,keV}/\nu_{2500\mbox{\rm\,\scriptsize\AA}})$, where
$f_{\rm 2\,keV}$ and $f_{2500\mbox{\rm~\scriptsize\AA}}$ are the flux
densities at 2\,keV and 2500\,\AA, respectively. The \aox\ values were
derived using the photon indices and fluxes in columns (2) and (3),
respectively, and the optical luminosities in column\,(6), assuming a
UV continuum of the form $f_{\nu}\propto \nu^{-0.5}$ (Vanden Berk \et
2001). The \aox\ values for our sources are consistent with the
expected values, given their optical luminosities (e.g., Steffen \et
2006).

\begin{figure*}
\centering
\includegraphics[width=17cm]{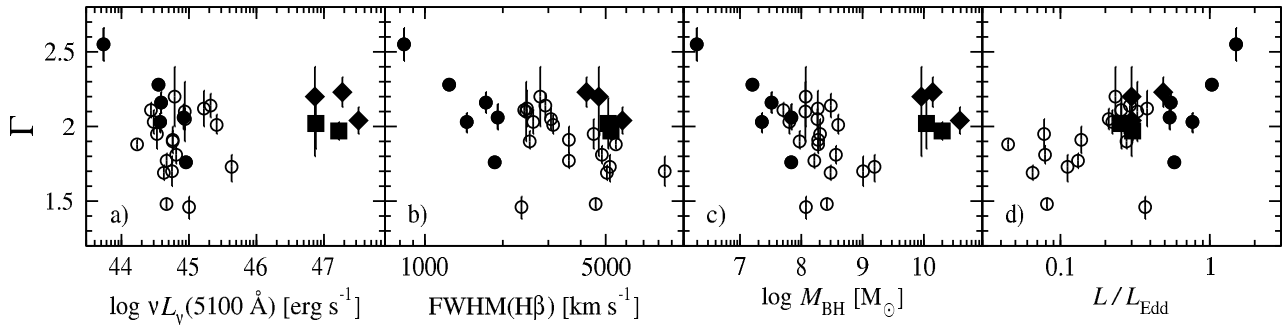}
\caption{The hard-\xray\ photon index vs. ({\it a}) $\nu
  L_{\nu}(5100\,\mbox{\AA})$, ({\it b}) FWHM(\hb), ({\it c}) \mbh, and
  ({\it d}) \lledd. Open circles mark PG quasars at $z<0.5$ from the
  P05 sample; the NLS1s in that sample are marked with filled
  circles. Diamonds mark luminous PG quasars at $z\sim1.5-2$ with
  archival \xmm\ data. \qthirteenlong\ and \hetwotwo\ are marked with
  squares.}
\label{GE}
\end{figure*}

\section{Is $\Gamma$ an Accretion-Rate Indicator?}
\label{indicator}

In Fig.\,\ref{GE} we plot $\Gamma$ vs.~$\nu
L_{\nu}(5100\,\mbox{\AA})$, FWHM(\hb),~\mbh, and \lledd\ for
\qthirteenlong\ and \hetwotwo. Fig.\,\ref{GE} also includes 28
unabsorbed Palomar Green (PG) RQQs (Schmidt \& Green 1983) which have
high-quality \xmm\ and optical data; 25 of the quasars are at $z<0.5$
and three, PG\,1247$+$267, PG\,1630$+$377, and PG\,1634$+$706, are at
\hbox{$z\sim1.5-2$}. Photon indices in the 2--12\,keV rest-frame band
for 27 of the PG quasars were obtained from Table\,3 of P05; the
photon index of PG\,1247$+$267 was obtained from Page \et
(2004b). Optical data for the PG quasars were obtained from Neugebauer
\et (1987), Boroson \& Green (1992), Nishihara \et (1997), and
McIntosh \et (1999), and processed with equations (\ref{eq:mbh}) and
(\ref{eq:lledd}). The fact that the \xray\ and optical data are not
contemporaneous may provide a source for scatter in
Fig.\,\ref{GE}. However, since most of the PG quasars do not exhibit
significant variations in $\Gamma$ (e.g., George \et 2000) and
\hb\ width (Kaspi \et 2000), and their optical-continuum flux
variations are typically at a level of \ltsim50\% over timescales of
several years (Kaspi \et 2000), the data in Fig.\,\ref{GE} are not
expected to be strongly affected by variability (see also equations
[\ref{eq:mbh}] and [\ref{eq:lledd}]).

We computed Spearman rank-correlation coefficients between $\Gamma$
and $\nu L_{\nu}(5100\,\mbox{\AA})$, FWHM(\hb), \mbh, and \lledd\ for
the 25 low-luminosity, $z<0.5$ sources. We found that, with the
exception of $\nu L_{\nu}(5100\,\mbox{\AA})$, the investigated
properties are significantly correlated with $\Gamma$ (with 99.9\%
confidence), in agreement with Porquet \et (2004) and P05. The
$\Gamma$-\mbh\, and $\Gamma$-\lledd\ correlations are largely a
consequence of the $\Gamma$-FWHM(\hb) correlation (Brandt \et 1997),
since both \mbh\ and \lledd\ depend strongly on FWHM(\hb), and the
sample spans a relatively narrow luminosity range [$\nu
  L_{\nu}(5100\,\mbox{\AA})\sim$10$^{44}$--$10^{46}$\,ergs\,s$^{-1}$;
  see eqs. (\ref{eq:mbh}) and (\ref{eq:lledd}), and Fig.\,\ref{GE}{\it
    a}]. The absence of a $\Gamma$-$L$ correlation has been observed
for larger AGN samples spanning broader luminosity ranges (e.g.,
Shemmer \et 2005; Vignali \et 2005).

We repeated the above correlations, adding the five luminous quasars
at $z\sim2$ to the analysis. We found that $\Gamma$ remains
uncorrelated with $\nu L_{\nu}(5100\,\mbox{\AA})$, and that the
$\Gamma$-\mbh\ correlation disappeared. This is a consequence of
extending the luminosity and \mbh\ ranges by $\sim3$ orders of
magnitude, while our measured $\Gamma$ values are typical of
high-redshift quasars (e.g., Shemmer \et 2005). The
$\Gamma$-FWHM(\hb) correlation remained significant with 99.9\%
confidence, but the (anti-)correlation coefficient dropped from 0.61
to 0.44. The $\Gamma$-\lledd\ correlation remained significant, with
99.9\% confidence, and the correlation coefficient, 0.60, has not
changed. This supports the hypothesis that the \xray\ photon index
depends primarily on \lledd.

To test the hypothesis that $\Gamma$ depends primarily on \lledd, we
considered AGNs from Fig.\,\ref{GE}{\it b} with
3500$<$FWHM(\hb)$<$6000\,\kms\ [which is the FWHM(\hb) interval of the
  five luminous $z\sim2$ sources as well as that of the majority of
  the S04 quasars], and checked the significance of the deviation
between the $\Gamma$ values of the five luminous sources and the
$\Gamma$ values of the eight $z<0.5$ PG quasars in that FWHM(\hb)
interval. A Mann-Whitney (MW) nonparametric rank test shows that the
$\Gamma$ distributions of the five high-luminosity quasars and the
eight moderate-luminosity quasars are significantly different with
99.9\% confidence. We also performed a KS test on the two $\Gamma$
distributions and found that the $\Gamma$ values of the two groups of
quasars cannot be drawn from a single distribution with 99\%
confidence. On the other hand, MW and KS tests showed that, in the
0.2\ltsim\lledd\ltsim0.5 range (representing the range of
\lledd\ values of the five luminous quasars as well as that of the
majority of the S04 sources), the $\Gamma$ values of the two groups of
quasars are not significantly different (as can be clearly seen from
Fig.\,\ref{GE}{\it d}).

Compton-reflection contributions to the \xray\ spectra of all 30
quasars are expected to be negligible given their moderate--high
luminosities (e.g., Page \et 2004a, but see also
Jim{\'e}nez-Bail{\'o}n \et 2005). However, Page \et (2004b) detected
Compton reflection in PG\,1247$+$267. We reanalyzed the
\xray\ spectrum of this source, and found that the hard-\xray\ excess
has no significant effect on $\Gamma$, whether fitted with or without
a Compton-reflection component (consistent with Page \et 2004b). We
also searched for systematic differences between the analyses of
\qthirteenlong\ and \hetwotwo\ and the P05 sources by reanalyzing the
\xmm\ spectra of PG\,1630$+$377, PG\,1634$+$706, and the eight $z<0.5$
PG quasars with 3500$<$FWHM(\hb)$<$6000\,\kms. We reproduced the P05
photon indices in the rest-frame \hbox{2--12\,keV} band with no
systematic deviations from their values. Additionally, fitting the
data of \qthirteenlong\ and \hetwotwo\ from rest-frame
\hbox{2--12\,keV} (observed-frame 0.6--3.5\,keV) did not alter our
results significantly. We note that for five $z<0.5$ PG quasars with
3500$<$FWHM(\hb)$<$6000\,\kms\ the $\Gamma$ values from P05 are
consistent with those from Porquet \et (2004), and are somewhat lower
than those from Brocksopp \et (2006), due perhaps to analyses
differences.

Our results suggest that $\Gamma$ depends primarily on \lledd, as
pointed out by e.g., Brandt \& Boller (1998) and Laor (2000). The key
difference between this work and previous studies (e.g., Porquet \et
2004; Wang, Watarai, \& Mineshige 2004; Bian 2005) is that we expanded
the quasar parameter space by adding highly luminous sources to the
correlations (see Fig.\,\ref{GE}{\it a}). Such luminous quasars are
found only at \hbox{$z\sim$1--4}, and their \hb\ lines, required for
\mbh\ and \lledd\ determinations (see e.g., S04; Baskin \& Laor 2005),
are shifted into the observed near-infrared band. \xray\ and
\hb\ spectral-region data of additional high-luminosity quasars are
crucial to break the degeneracy between the $\Gamma$-FWHM(\hb) and
\hbox{$\Gamma$-\lledd} correlations and conclude that the
\xray\ power-law photon index depends on the accretion rate.
Ultimately, a large enough inventory of \xray\ and
\hb\ spectral-region data for luminous, high-redshift quasars will
allow testing of the hypothesis that such sources are analogous to
NLS1s (e.g., Grupe \et 2006).

\acknowledgments

This work is based on observations obtained with \xmm, an ESA science
mission with instruments and contributions directly funded by ESA
Member States and the USA (NASA). We thank an anonymous referee for
constructive comments, and George Chartas and Aaron Steffen for
fruitful discussions. We gratefully acknowledge the financial support
of NASA grant \hbox{NNG05GP00G} (O.\,S, W.\,N.\,B), NASA LTSA grant
\hbox{NAG5-13035} (O.\,S, W.\,N.\,B), and the Zeff Fellowship at the
Technion (S.\,K.). This work is supported by the Israel Science
Foundation grant 232/03.

\end{document}